# LEAGUE CHAMPIONSHIP ALGORITHM (LCA) BASED JOB SCHEDULING SCHEME FOR INFRASTRUCTURE AS A SERVICE CLOUD

Shafi'i M. Abdulhamid*[1] and Muhammad S. Abd Latiff[2]

[1, 2] Faculty of Computing, Universiti Teknologi Malaysia, Johor Bahru MALAYSIA.
[1]Department of Cyber Security Science, Federal University of Technology Minna, NIGERIA.
(E-mail: shafii.abdulhamid@futminna.edu.ng, shafie@utm.my)

## INTRODUCTION

League Championship Algorithm (LCA) is a population based algorithmic framework for global optimization over a continuous search space first proposed by Kashan (2009). It is a Swam optimization algorithm. A general characteristic between all population based optimization algorithms similar to the LCA is that they both try to progress a population of achievable solutions to potential areas of the search space when seeking the optimization. LCA is a newly proposed stochastic population based algorithm for continuous global optimization which tries to imitate a championship situation where synthetic football clubs participate in an artificial league for a number of weeks. This algorithm has been tested in many areas and performed creditably well as compared to other known optimization schemes or heuristics algorithms.

In this research work, we proposed a job scheduling algorithm based on the LCA optimization technique for the IaaS cloud. Two other well established algorithms i.e. First Come First Served (FCFS) and Last Job First (LJF) were used to evaluate the performance of the proposed algorithm.

The parameters used for measuring the scheduling algorithms in this experiment are based on three factors - Average Response Time, Makespan Time, and the Average Completion Time. The data set was formed by using workload 5000 processes (jobs). The experiment was performed by varying the number of virtual machines in the IaaS cloud from 10 to 130. All three algorithms assumed to be non-preemptive.

**MAIN RESULTS**

Figure 1 below shows that the average total completion times as calculated by the three scheduling schemes for the 5000 jobs. The average total completion time as processed by the LCA scheduling algorithm is shorter than the other two algorithms, i.e. LJF and FCFS. The LJF has the longest completion time amoung the algorithms under consideration in this experiment.

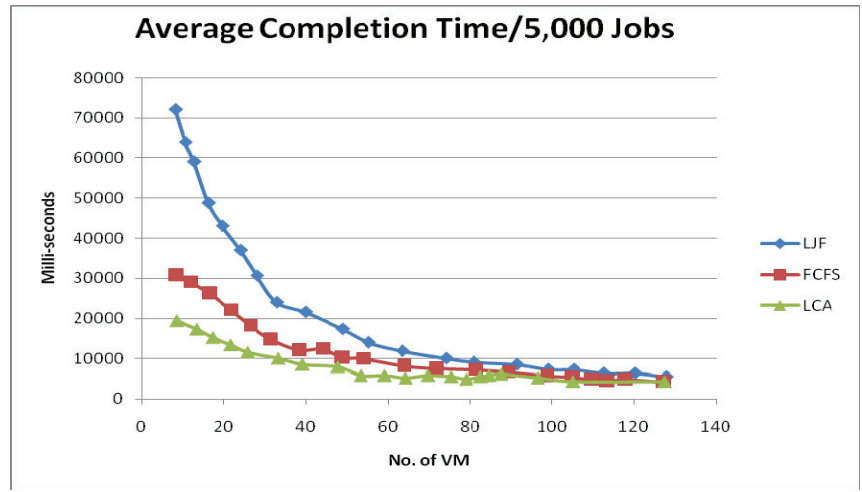

**Figure 1.** Average Completion Time for 5,000 Jobs

The results obtained from the infrastructure as a service (IaaS) cloud environment also shows that, LCA scheduling algorithm perform better than the FCFS and the LJF algorithms. Especially in reducing the average response time, the average completion time (Figure 1 above) and the makespan time of the scheduling algorithms.

**Acknowledgment:** The first author would like to express his appreciation for the support of UTM International Doctoral Fellowship (IDF).